\documentstyle[prl,aps,twocolumn]{revtex}\begin{document}\iftrue 

\draft
\title{ \hfill {\small\rm SMC-PHYS-168}\\
Brane Induced Gravity from its Dynamical Origin }

\author{		Keiichi Akama\footnote{E-mail: akama@saitama-med.ac.jp }}
\address{	Department of Physics, Saitama Medical College,
		Kawakado, Moroyama, Saitama, 350-04, Japan}
\maketitle
\begin{abstract}
We develop a quantum theoretical formalism 
	to derive the brane gravity 
	from its origin --- brane-generating dynamics in higher dimensions.
Based on it, we discuss characteristic properties of 
	the brane induced gravity, as well as brane induced field theory.
\end{abstract}

\vskip15pt
\else
\draft
\twocolumn[
\widetext

\hfill SMC-PHYS-168

\hfill hep-th/0304@@@

\centerline{\large\bf	Brane Induced Gravity from its Origin } 

\vskip 10pt
\centerline{     Keiichi Akama }
\centerline{\small\sl       Department of Physics, Saitama Medical College,
                Kawakado, Moroyama, Saitama, 350-0496, Japan}
\centerline{\small   e-mail: \tt akama@saitama-med.ac.jp }
\vskip 15pt

\leftskip 15mm\rightskip 15mm
\begin{abstract}
\baselineskip  = 10pt
We develop a quantum theoretical formalism 
	to derive the brane gravity 
	from its origin --- brane-generating dynamics in higher dimensions.
Based on it, we discuss characteristic properties of 
	the brane induced gravity, as well as brane induced field theory.
\end{abstract}

\vskip15pt
]

\narrowtext
\fi

The idea of the induced gravity was proposed by Sakharov in 1967,
	where the Einstein gravity is identified with  
	the quantum fluctuations of metric elasticity of 
	space \cite{Sakharov}.
The field theoretical formulations of the induced gravity 
	were developed about a decade later 
	\cite{Sakharov2,ACMTetc,A78,AdlerZeeAFV}, 
	inspired by the progresses in the unified theory \cite{TCAetc}
	with induced (composite) gauge fields \cite{Bjorken}
	and by those in the loop diagram calculations \cite{QG} 
	in the quantum gravity.
The spacetime metric or vielbein is a quantum composite object
	which is absent in the microscopic description \cite{A78}.
The fundamental action for the scalar-induced gravity is nothing
	but the Nambu-Goto action \cite{NambuGoto} of the brane, 
	if we identify the fundamental scalar fields 
	with the space-time coordinates.
This fact naturally leads us to the idea 
	that we live on a dynamically generated braneworld
	with induced gravity \cite{A83,BW}.
Besides the gravitational field, 
	the extrinsic curvature and
	the normal-connection gauge field 
	become dynamical \cite{A87etc}. 
In this view the Einstein gravity on the brane
	is induced also from the extrinsic curvature effects 
	via the Gauss-Codazzi-Ricci equation \cite{A88,SMS}.
Recently, the brane induced gravity attracted revived attentions 
	with interesting developments and some disputes \cite{BIG}.
In its derivation, however, 
	the quantum nature of the system has so far been obscure, 
	in spite of its essential reliance on the quantum fluctuation effects.
In the present work, 
	we would like to develop a fully quantum theoretical formalism 
	to derive the Einstein-like gravity on the brane 
	from the origin of brane-generating dynamics in higher dimensions.

Consider a set of fields $\{\Phi\}$
	(in general, including higher spin and spinor fields)
	in $p+q+1$-space-time with the action
\begin{equation}
S=\int L(\{\Phi\},\{\partial_K\Phi\})d_{p+q+1} X,
		\label{Sflat}
\end{equation} 
	which is invariant under translation, Lorentz transformation, 
	and internal symmetry $G$ 
	with degenerate potential minima.
Let us assume that the equation of motion from (\ref{Sflat})
	has a static soliton solution 
	$ \Phi=\Phi_{\rm sol}(X^{\underline a})$
	localized around the $p$-brane $X_{\underline a }=0$ \cite{suffix}, 
	such that 
\begin{equation}
\Phi_{\rm sol}\sim\Phi_{\rm vac}+O(e^{-r/\delta})\ \ {\rm for}\ \ 
	r=\sqrt{-X^{\underline a} X_{\underline a }}\rightarrow \infty,
	\label{Phisol}
\end{equation}  
	where $\delta $ is a constant, 
	and $ \Phi^{\rm vac} $ is a 
	singular function
	which takes the values of the degenerate potential minima 
	except at the singularity at $ X^{\underline a }=0$. 
This is interpreted as a flat p-brane localized around $ X^{\underline a }=0$.
For example, the kink solution of the scalar field in the double-well potential,
	the Nielsen-Olessen vortex in U(1) gauge-Higgs model, 
	and the Skyrmion with non-vanishing pseudo-scalar mass 
	are known to have these properties.
Learning from the existing examples, we assume that 
	the solution have ``spherical symmetry" of the type
\begin{equation}
	\forall {\cal R}\in{\rm O}(q), \exists g\in G, 
	U({\cal R})V(g)\Phi_{\rm sol}=\Phi_{\rm sol}
	\label{spherical}
\end{equation} 
where $U({\cal R})$ and $V(g)$ are representation matrices 
	of the rotation group O($q$) 
	and the internal symmetry group $G$, respectively.

The quantum dynamics of the system is described by the generating functional
	of the Green functions: 
\begin{equation}
W(J)=\int d\Phi \exp\left[iS+iJ*a(\Phi)\right],
	\label{WJ=}
\end{equation}   
where $J$ is a function of $X^K$,
	$a(\Phi)$ is a function of $\Phi$,
	the asterisk ``*" indicates convolution,
	and the path-integration measure $d\Phi$ 
	is normalized by $W(0)=1$.
We consider the sector of configurations connected with $\Phi_{\rm sol}$
	without high energy barriers.
The configurations of $\Phi_{\rm sol}$ and 
	its small fluctuations would give dominant contributions
	to the path integral.
Among the fluctuations, the translation zero modes 
	give rise to shifts of the brane position.
The shifts can, in general, depends on $X^a$, which means that
	the brane becomes curved.
We should also take into account the globally large fluctuations
	as far as the local curvature is small, 
	since they could have comparable contributions.
We denote the position of the brane by {\boldmath $y \it(x^\mu)$}
	with $p+1$ parameters $x^\mu$.
We define the frame of orthonormal tangent vectors {\boldmath $n\it_k(x^\mu)$}, 
	and normal vectors {\boldmath $n\it_{\underline k}(x^\mu)$}
	at each point on the brane.
Then the derivatives {\boldmath $y$}$_{,\mu}^{\ }$ 
	and {\boldmath $n$}$_{I,\mu}^{\ }$
	are expanded in terms of {\boldmath $n\it_{I}$} \cite{suffix}:
\begin{equation}
	\mbox{\boldmath $y$}_{,\mu}^{\ } 
	= \mbox{\boldmath $n$}_k^{\ } e^k _{\ \mu},\ \ \
	\mbox{\boldmath $n$}_{I,\mu}^{\ } 
	= \mbox{\boldmath $n$}_K^{\ } \omega^K _{\ I\mu},\ \ \
\end{equation}  
	where $e^k _{\ \mu}$ is the vielbein on the brane, 
	and $\omega^K _{\ I\mu}$ is the connection coefficient.
Among them, 
	$\omega^k _{\ i\mu}$ is the tangential connection,
	$\omega^{\underline k} _{\ \mu\nu}$ is the extrinsic curvature,
	and $\omega^{\underline k} _{\ {\underline i}\mu}$
	is the normal connection of the brane.

Then we introduce the curvilinear coordinate $x^M=(x^\mu,x^{\underline m})$ by
\begin{equation}
	\mbox{\boldmath $X$} = \mbox{\boldmath $y$}(x^\mu)
	+ x^{\underline m} \mbox{\boldmath $n$}_{\underline m}(x^\mu)
	\label{X=}
\end{equation}
The coordinate should be defined otherwise appropriately 
	in the far apart regions 
	where the $ x^{\underline m}$ coordinates by (\ref{X=}) intersect
	each other.
We define the field $\phi$ in the curvilinear coordinate $x^M$ 
	and the orthonormal frame {\boldmath $n$}$_I$ by
\begin{equation}
	\Phi(X^K)=U(L)V(g)\phi(x^M), \label{def_phi}
\end{equation}  
where $L\in$O$(p+q,1)$ is defined by $L_I^{\ K}=(${\boldmath$n$}$^K)_I$, 
	and $g\in G$ is 	the element which exists by (\ref{spherical})
	with $L_{\underline i}^{\ \underline k}
	={\cal R}_{\underline i}^{\ \underline k}$.
The derivative of $\Phi$ is transformed as follows \cite{A87etc}
\begin{equation}
	\mbox{\boldmath$\partial$}
	\Phi(X^K)=U(L)V(g) \mbox{\boldmath$n$}^K D_K\phi(x^M) 
\label{del Phi=}
\end{equation}   
	with $ D_{\underline k}=\partial_{\underline k} $ and
\begin{eqnarray} &&
	D_k=(K^{-1})_k^{\ l} e_l^{\ \mu}
	\Big\{\partial_\mu
	-{i\over 2}\omega_{ij\mu}\sigma^{ij}
\label{Dk=}
\cr&&\ \ \ \ 
	-i\omega_{{\underline i}j\mu}\sigma^{{\underline i}j}
	-{i\over 2}\omega_{\underline {ij}\mu}
	(\sigma^{\underline {ij}}+\lambda^{\underline {ij}}
	+2ix^{\underline i}\partial^{\underline j})
	\Big \},
\end{eqnarray}
	where 
	$K_k^{\ l}=\eta_k^{\ l}-x^{\underline i}\omega_{\underline ik}^{\ \ l}$,
	and
	$\sigma^{IJ}$ and $\lambda^{\underline {ij}}$
	are the representation matrices of the group generators of
	O($p+q,1$) and $G$, respectively.
Then the action is rewritten as
\begin{equation}
	S=\int e\det K_k^{\ l} 
	L(\{\phi\},\{D_K^{\ }\phi\})\Pi dx^{\mu}\Pi dx^{\underline {k}}
	\label{S=2}
\end{equation}
	with $e=\det e^k_{\ \mu}$.

By (\ref{spherical}) and (\ref{def_phi}), 
	$\phi=\Phi_{\rm sol}( x^{\underline k})$
	satisfies the equation of motion 
	except in the far-apart singular region of (\ref{X=}).
The errors caused by them are exponentially suppressed 
	for small brane-curvatures.
We neglect the errors, and expand $\phi$ in terms of
	the complete set $\{\varphi_n(x^{\underline k})\}_{n=0,1,2,\cdots}$
	of the fluctuations of the soliton $\Phi_{\rm sol}$:
\begin{equation}
	\phi(x^\mu,x^{\underline k})=\Phi_{\rm sol}( x^{\underline k})
	+\sum_n\xi_n(x^\mu)\varphi_n(x^{\underline k}),
\label{expand}
\end{equation}
where $\xi_n(x^\mu)$ is the appropriate coefficient depending on $x^\mu$.
We change the path-integration variable in $W(J)$ in (\ref{WJ=}) 
	from $\Phi(X^M)$ to $\xi_n(x^\mu)$. 
Since (\ref{expand}) is linear in them,  
	the Jacobian is a numerical constant 
	which is absorbed by path-integral measure normalization.

The complete set $\{\varphi_n(x^{\underline k})\}$ include 
	the zero modes 
	$\varphi_{\rm Tr0}^{\underline k}(x^{\underline k})$ and
	$\varphi_{\rm Lz0}^{\underline k l}(x^{\underline k})$
	associated with the translation 
	and the Lorentz invariance, respectively,
	which are broken by the brane.
They cause shifts in the brane position {\boldmath$y$}$_\Phi^{\ }(x^\lambda)$
	and orthonormal frame {\boldmath$n$}$_\Phi^{K}(x^\lambda)$,
	and cause double counting of the field configurations. 
We choose {\boldmath$y$}$_\Phi^{\ }(x^\lambda)$ 
	and {\boldmath$n$}$_\Phi^{K}(x^\lambda)$
	so that the coefficients $\xi^{\rm Tr0}_{\underline k}(x^\lambda)$
	of $\varphi_{\rm Tr0}^{\underline k}(x^{\underline k})$ 
	and those $\xi^{\rm Lz0}_{\underline k l}(x^\lambda)$ of
	$\varphi_{\rm Lz0}^{\underline k l}(x^{\underline k})$ vanish.
These choice prescriptions work well 
	for the configurations with large contributions, i.e.\ 
	for those with small brane curvatures.

Then we insert into the path-integral in (\ref{WJ=}) 
\begin{eqnarray}&&
	1=\int d\mbox{\boldmath$y$}
	\delta (\mbox{\boldmath$y$}
	-\mbox{\boldmath$y$}_\Phi^{\ }  )
	d\mbox{\boldmath$n$}_I^{\ } 
	\delta (\mbox{\boldmath$n$}_I^{\ } 
	-\mbox{\boldmath$n$}_{\Phi I}^{\ })
\label{1=dy}
\\&&
	1=\int de_{k\mu}^{\ } 
	\delta (e_{k\mu}^{\ } 
	-\mbox{\boldmath$n$}_k^{\ }  
	\mbox{\boldmath$y$}_{,\mu}^{\ }) 
	d\omega_{IJ\mu}^{\ }
	\delta (\omega_{IJ\mu}^{\ } 
	-\mbox{\boldmath$n$}_I^{\ } 
 	\mbox{\boldmath$n$}_{J,\mu}^{\ }) 
\label{1=de}
\end{eqnarray}
where integrations and delta functions are product over $x^\mu$
	and over independent tensorial components.
We identify {\boldmath$y$} and {\boldmath$n$}$_I$
	with the wrongly chosen brane position and orthonormal frame,
	respectively, in the above-prescribed choice of 
	{\boldmath$y$}$_\Phi^{\ }$ and {\boldmath$n$}$_{\Phi I}^{\ }$.
Then the above choice prescription leads to
\begin{eqnarray}&&
	\delta (\mbox{\boldmath$y$}
	-\mbox{\boldmath$y$}_\Phi^{\ })
	=\delta (\xi_{\underline k}^{\rm Tr0})\delta(0)^{p+1}
\\&&
	\delta (\mbox{\boldmath$n$}_I
	-\mbox{\boldmath$n$}_{\Phi I}^{\ })
	=\delta (\xi_{\underline k l}^{\rm Lz0})
	\delta(0)^{p(p+1)/2+q(q-1)/2}.
\end{eqnarray}   
We regularize the infinities of $\delta(0)$ by the definition
\begin{eqnarray}
	\delta(z)=\lim_{\epsilon\rightarrow0}\delta_\epsilon(z),\ \ \ \
	\delta_\epsilon(z)=\int_{-1/\epsilon}^{1/\epsilon} e^{ikz}dk, 
\end{eqnarray}
	where the limit is taken at the final stage of calculation 
	(though $\epsilon$ is not explicitly shown hereafter).
Note that the Lorentz invariance 
	is restored in the limit.

Now we prove the general path-integral formula for functions 
	$f (x^\lambda)$ and $ v_\mu(x^\lambda)$
\begin{eqnarray}&&
	\int df \delta\ (v_\mu -  f_{,\mu})
	= c\delta (v_{[\mu,\nu]}) 
\label{formula}
\end{eqnarray} 
	where $c$ is an overall constant factor \cite{suffix}.
We discretize the $x^\mu$-space 
	into $N_0\times N_1\times \cdots\times N_p$ lattice
	with the lattice constant $h$ by
\begin{eqnarray}&&
	{\cal L}= \{\hat x= k_0 \hat h_0+ k_1 \hat h_1
	\cdots+ k_p \hat h_p
\cr&&\ \ \ \ \ \ \ \ \ \ \ \ \ \ \ 
	| k_\mu=1,2,\cdots,N_\mu(\mu=0,1,\cdots,p) \}
\label{lattice}
\end{eqnarray}
where the hat indicates a $p+1$-dimensional vector, and 
	$\hat x$ and $\hat h_\mu$ are defined by 
	$(\hat x) ^\mu =x^\mu$, and 
	$(\hat h_\mu)^\nu=h\delta_\mu^\nu$.
Then, the left-hand side of the (\ref{formula})
	is the $h\rightarrow 0$ limit of 
\begin{eqnarray}&&
	\prod_{\hat x\in {\cal L}}\left[\int df(\hat x)
		\prod_{\mu=0}^p\delta\left(v_\mu(\hat x)
	-{f (\hat x)- f (\hat x-\hat h_\mu) \over h} \ \right)
	\right]
	\nonumber
\\&&\ \ \ \label{f1}
\end{eqnarray}
We rewrite the delta function with $\mu=0$ in (\ref{f1}) as
\begin{eqnarray}&&
	h\delta\left(hv_0(\hat x)
	-f (\hat x)+ f (\hat x-\hat h_0) \ \right).
\label{delta1}
\end{eqnarray}
and rewrite those with $\mu=1,2,\cdots,p$ as
\begin{eqnarray}&&
	\delta\left(v_\mu(\hat x)-v_\mu(\hat x-\hat h_0)
	-v_0(\hat x)+v_0(\hat x-\hat h_\mu)\right)
\label{delta2}
\end{eqnarray}
using the expressions for $ v_\mu(\hat x-\hat h_0)$,
	$v_0(\hat x)$ and $v_0(\hat x-\hat h_\mu)$
	from their respective delta functions.
To restore symmetry under permutations of $x_\mu$ in (\ref{delta2}),
	we multiply them by the factors
\begin{eqnarray}&&
	\delta\left(v_\mu(\hat x)-v_\mu(\hat x-\hat h_\nu)
	-v_\nu(\hat x)+v_\nu(\hat x-\hat h_\mu)\right)
\label{delta3}
\end{eqnarray}
with $0<\mu<\nu\le p$. 
Owing to (\ref{delta2}), the expression (\ref{delta3}) 
	is equal to an infinite constant $\delta(0)$. 
Since there are no delta functions at the boundary $x^\nu=0$, 
	we need to impose appropriate boundary conditions
	for the functions.
Using (\ref{delta1})--(\ref{delta3}), we rewrite (\ref{f1}) as 
\begin{eqnarray}&&
	\prod_{\hat x\in {\cal L}}\Bigg[\int df(\hat x)
	\delta\left(hv_0(\hat x)
	-f (\hat x)+ f (\hat x-\hat h_0) \ \right)	
\cr&&
	\prod_{\mu<\nu}
	\delta\left({{v_\mu(\hat x)-v_\mu(\hat x-\hat h_\nu)}\over h}
	-{{v_\nu(\hat x)-v_\nu(\hat x-\hat h_\mu) }\over h}\right) \Bigg]
	\nonumber
\\&&\ \ \ \ / h^{p(p+1)/2-1}\delta(0)^{ p(p-1) /2}\label{f2}
\end{eqnarray}
We perform path-integration with respect to $f$, 
	and take the limit $h\rightarrow 0$, 
	which gives the right-hand side of (\ref{formula})
	with $c= h^{p(p+1)/2-1}\delta(0)^{ p(p-1) /2}\label{f2}$. 
	{\it q.e.d.}

We apply (\ref{formula}) to ingredients of (\ref{1=dy}) and (\ref{1=de}): 
\begin{eqnarray}&&
	\int d\mbox{\boldmath$y$} 
	\delta (e_{k\mu}^{\ } 
	-\mbox{\boldmath$n$}_k^{\ } 
	\mbox{\boldmath$y$}_{,\mu}^{\ }) 
	=c_1
	\delta (\omega^k_{\ [\mu\nu]} 
	+e^k_{\ [\mu,\nu]}) 
\label{delta(omega-e)}
\\&&
	\int d\mbox{\boldmath$n$}_I^{\ }
	\delta (\omega_{IJ\mu}^{\ } 
	-\mbox{\boldmath$n$}_I^{\ }\mbox{\boldmath$n$}_{J,\mu}^{\ }) 
\cr&&\ \ \ \ \ \ \ \ \ \ \ \ \ \ \ \ \ \ \ \ 
	=c_2
	\delta (\omega_{IJ{[\mu,\nu]}}^{\ }
	+\omega _{KI[\mu}^{\ } \omega ^K_{\ J\nu]}) 
\label{delta(GCR)}
\end{eqnarray}
where $c_1$ and $c_2$ are constants. 
In deriving (\ref{delta(omega-e)}) and (\ref{delta(GCR)}),
	we have used the fact that $\det[(${\boldmath$n$}$_I)_J]=1$.
The expression (\ref{delta(omega-e)}) imposes the usual relation
	between $\omega_{ij\mu}$ and $e_{k\mu}$:
\begin{eqnarray}
	\omega_{ij\mu}^{\ }=(c_{[ij]\mu}^{\ }-c_{\mu ij}^{\ })/2{\rm\ with\ }
	c_{i\mu\nu}^{\ }=e_{i[\mu,\nu] ^{\ }},
\label{omega=}
\end{eqnarray}
while that in (\ref{delta(GCR)}) imposes 
	the Gauss-Codazzi-Ricci equation on $\omega_{IJ\mu}^{\ }$.
Collecting the results, we obtain
\begin{eqnarray}&&
	W(J)=\int de_{k\mu}
	d\omega_{\underline i j\mu}
	d\omega_{\underline{ij}\mu}
	{ \Pi}_n d\xi_n
\cr&&\ \ \ \ \ \ \ \ \ \ \ \ \ \ \ \ \ 
	\delta (\omega_{IJ{[\mu,\nu]}}^{\ }
	+\omega _{KI[\mu}^{\ } \omega ^K_{\ J\nu]}) 
	e^{iS+iJ*a(\Phi)},
\end{eqnarray}
where  ${ \Pi}_n d\xi_n $ does not include the zero modes,
	and $\omega_{ij\mu}$ are taken as expressed 
	in terms of $e_{k\mu}$ by (\ref{omega=}).

Now we expand $S$ in (\ref{S=2})
	in terms of $\varphi_n(x^{\underline k})$ in (\ref{expand}).
Among $\varphi_n(x^{\underline k})$, 
	those below energy threshold are localized around the brane.
We call the part of $S$ with only the localized modes as $S_{\rm loc}$,
	and the rest as $S_{\rm bulk}$.
We perform the $x^{\underline k}$ integration in $S_{\rm loc}$, 
	which gives the brane action $S_{\rm br}$.
The action $S_{\rm br}$ is a functional of 
	the vielbein $e_k^\mu$, 
	the extrinsic curvature $\omega_{\underline k \mu\nu}$,
	the coefficients $\xi_n^{\rm loc}$ of the localized modes,
	and their covariant derivatives ${\cal D}_\mu\xi_n^{\rm loc}$
	with respect to the gauge group $O(p,1)\times O(q)$,
	which are written in terms of the connection (\ref{omega=}) and 
	the normal-connection gauge field $\omega_{\underline{kl}\mu}$.

Then we expand $S_{\rm br}$ in terms of 
	$\omega_{\underline k \mu\nu}$, $\omega_{\underline{kl}\mu}$,
	$\xi_n^{\rm loc}$, and $\partial_\mu\xi_n^{\rm loc}$.
The coefficient of each term is calculable 
	if the model is specified (See e.g.\ \cite{A88}).
We expect that the lower order terms are important at low energies.
Since $\xi_n$ are the small fluctuations of the solution $\Phi_{\rm sol}$,
	the terms with single $\xi_n$ are absent in $S_{\rm br}$.
The lowest order terms are those bilinear in them,
	giving the mass and kinetic terms of $\xi_n^{\rm loc}$.
The higher terms including $\xi_n^{\rm loc}$ give their interaction terms.
Thus the field theory of $\xi_n^{\rm loc}$ is induced on the brane.

The normal connection gauge field $\omega_{\underline{kl}\mu}$
	appear only through the covariant derivative ${\cal D}_\mu\xi_n^{\rm loc}$,
	and has no mass and kinetic terms. 
The extrinsic curvature $\omega_{\underline k \mu\nu}$
	appear in the overall factor $\det K_k^{\ l} $,
	and twice in the in $D_k$ in(\ref{Dk=}). 
Terms solely with $\omega_{\underline k \mu\nu}$ come from 
\begin{eqnarray}&&
	\int e\det K_k^{\ l} 
	L(\{\Phi_{\rm sol}\},\{\partial_{\underline k}^{\ }\Phi_{\rm sol}\})
	\Pi dx^{\mu}\Pi dx^{\underline {k}},
	\label{S=3}
\\&&
	\det K_k^{\ l}=1-x^{\underline k}\omega_{\underline k \mu}^{\ \ \mu}
\cr&& \ \ \ \ \ \ \ \ 
	-{1\over2} x^{\underline k} x^{\underline l}
	(\omega_{\underline k \mu}^{\ \ \mu}\omega_{\underline l \nu}^{\ \ \nu}
	-\omega_{\underline k \mu}^{\ \ \nu}\omega_{\underline l \nu}^{\ \ \mu})
	+\cdots.
\end{eqnarray}
If we assume the parity invariance of the Lagrangian $L$,
	$\det K_k^{\ l} $ in (\ref{S=3}) reduces to
\begin{eqnarray}&&
	\det K_k^{\ l}=1 
	-{1\over2q} x_{\underline k} x^{\underline k}
	(\omega_{\underline l \mu}^{\ \ \mu}\omega_{\ \nu}^{\underline l \ \nu}
	-\omega_{\underline l \mu}^{\ \ \nu}\omega_{\  \nu}^{\underline l \ \mu})
	+\cdots.
\label{detK=}
\end{eqnarray}
The second term of (\ref{detK=})
	give rise to the mass term of $\omega_{\underline k \mu\nu}$,
	but no kinetic term exists.
The vielbein $e_{k\mu}$ appear in the overall factor $e$, 
	in the connection (\ref{omega=}) 
	in the covariant derivative ${\cal D}_\mu\xi_n^{\rm loc}$, and 
	various places to make each term reparametrization invariant.
It solely appear only in the term proportional to $e$ in (\ref{S=3})
	(which is the cosmological term), 
	and no kinetic term of $e_{k\mu}$ exists apparently.

The Gauss equation from (\ref{delta(GCR)}), however, converts 
	the $\omega_{\underline k \mu\nu}$ mass term in (\ref{S=3})
	into the Einstein-Hilbert action, 
	and $e_{k\mu}$ becomes dynamical.
Furthermore, the quantum fluctuations of $\xi_n^{\rm loc}$
	give rise to the kinetic terms of $e_{k\mu}$,
	$\omega_{\underline k \mu\nu}$, and $\omega_{\underline{kl}\mu}$,
	making them dynamical, 
	as has been extensively studied 
	\cite{Sakharov,Sakharov2,ACMTetc,A78,AdlerZeeAFV,TCAetc,Bjorken}.  
Naively they are divergent in the ultraviolet region,
	and usually some cutoff is assumed.
Such a cutoff is rather natural in the braneworld,
	since the brane descriptions themselves would 
	no longer be appropriate
	at very high energies, or very short distances
	compared with characteristic scales of the brane width and depth.
Phenomenologically, the cutoff should be of the order of the Planck scale.
Because of the large cutoff scale,
	the Einstein-Hilbert action dominates over 
	the other kinetic terms of $e_{k\mu}$
	(e.g.\ $R^2$ terms) at low energies.
The same effects also induce a cosmological term with the large scale,
	and, phenomenologically, should be fine-tuned 
	so that it cancels that in (\ref{S=3}).
The extrinsic curvature $\omega_{\underline k \mu\nu}$ 
	acquires masses of the order of the Planck scale,
	since no symmetry protect them.
The gauge field $\omega_{\underline{kl}\mu}$ would acquire mass
	if some of the $\varphi_n^{\rm loc}$ violate $O(q)$ symmetry.
The gravity is Einstein-like, but a difference is that $e_{k\mu}$, 
	$\omega_{\underline k \mu\nu}$, and $\omega_{\underline{kl}\mu}$ 
	also obey the Gauss-Codazzi-Ricci equation. 
Quantum fluctuations due to the bulk action $S_{\rm bulk}$ 
	could also have some contributions
	to the kinetic terms of  $e_{k\mu}$,
	$\omega_{\underline k \mu\nu}$, and $\omega_{\underline{kl}\mu}$.
It would, however, be a difficult task to estimate them.

In conclusion, Einstein-like gravity with matter fields
	is induced on the dynamically generated braneworld.
For example, for two matter sources on the brane, 
	it automatically leads to the usual gravitational interaction 
	with spin-2 graviton exchange. 
All the features of the Einstein gravity are reproduced 
	at low energies on the brane.
The effective theory is derived from its origin in a fully quantum theoretical manner.
We expect that such origin-based studies
	would provide comprehensive answers to the disputes on the braneworlds.
It would be desired 
	to apply the formalism to various models of solitons,
	and to seek for the realistic models of fundamental fields
	as well as realistic cosmology.

The author would like to thank Professor T.\ Inami, Dr.\ T.\ Hattori,
	and Dr.\ H.\ Mukaida for discussions.
This work is supported by Grant-in-Aid for Scientific Research, 
	Japan Society for the Promotion of Science.


\begin{thebibliography}{99}
\bibitem {suffix}
Capital indices stand for $0,1,\cdots,p+q$, and
	lower case indices without (with) a underline
	stand for $0,1,\cdots,p $ ($p+1,\cdots,p+q $). 
$F_{,\mu}$ stand for $\partial_\mu F$.
The Greek (Latin) indices denote 
	the general coordinate (local Lorentz) indices.
The indices are raised, lowered and converted between Greek and Latin
	as usual.
The suffices with left and right pieces of
	square brackets indicate anti-symmetrization among them.
\end{thebibliography}
\end{document}